\documentclass[aps,pra,amsmath,superscriptaddress,amssymb,reprint]{revtex4-1}

\usepackage{graphicx}
\usepackage{amsmath}

\newcommand{\PhysicalUnit}[1]{\,\mathrm{#1}}

\newcommand{\MicroWatt}[0]{\PhysicalUnit{\mu W}}
\newcommand{\NanoSeconds}[0]{\PhysicalUnit{ns}}
\newcommand{\MilliSeconds}[0]{\PhysicalUnit{ms}}
\newcommand{\NanoMeter}[0]{\PhysicalUnit{nm}}
\newcommand{\MegaBarn}[0]{\PhysicalUnit{Mb}}
\newcommand{\MicroMeter}[0]{\PhysicalUnit{\mu m}}

\newcommand{\Seconds}[0]{\PhysicalUnit{s}}

\begin{document}

\title{Resonant photo-ionization of $\mathrm{Yb^+}$ to
  $\mathrm{Yb^{2+}}$} 

\author{Simon Heugel}
\affiliation{Max Planck Institute for the Science of Light,
  Guenther-Scharowsky-Str. 1/ building 24, 91058 Erlangen, Germany}
\affiliation{Friedrich-Alexander-Universit\"at Erlangen-N\"urnberg (FAU),
  Department of Physics, Staudtstr. 7/B2, 91058 Erlangen, Germany}

\author{Martin Fischer}
\affiliation{Max Planck Institute for the Science of Light,
  Guenther-Scharowsky-Str. 1/ building 24, 91058 Erlangen, Germany}
\affiliation{Friedrich-Alexander-Universit\"at Erlangen-N\"urnberg (FAU),
  Department of Physics, Staudtstr. 7/B2, 91058 Erlangen, Germany}

\author{Vladimir Elman}
\affiliation{Max Planck Institute for the Science of Light,
  Guenther-Scharowsky-Str. 1/ building 24, 91058 Erlangen, Germany}

\author{Robert Maiwald}
\altaffiliation[Present address: ]{Physikalisches Institut, 
University of Bonn, Wegelerstrasse 8, 53115 Bonn, Germany}
\affiliation{Max Planck Institute for the Science of Light,
  Guenther-Scharowsky-Str. 1/ building 24, 91058 Erlangen, Germany}
\affiliation{Friedrich-Alexander-Universit\"at Erlangen-N\"urnberg (FAU),
  Department of Physics, Staudtstr. 7/B2, 91058 Erlangen, Germany}

\author{Markus Sondermann}
\affiliation{Max Planck Institute for the Science of Light,
  Guenther-Scharowsky-Str. 1/ building 24, 91058 Erlangen, Germany}
\affiliation{Friedrich-Alexander-Universit\"at Erlangen-N\"urnberg (FAU),
  Department of Physics, Staudtstr. 7/B2, 91058 Erlangen, Germany}

\author{Gerd Leuchs}
\email[]{gerd.leuchs@mpl.mpg.de}
\affiliation{Max Planck Institute for the Science of Light,
  Guenther-Scharowsky-Str. 1/ building 24, 91058 Erlangen, Germany}
\affiliation{Friedrich-Alexander-Universit\"at Erlangen-N\"urnberg (FAU),
  Department of Physics, Staudtstr. 7/B2, 91058 Erlangen, Germany}
\affiliation{Department of Physics, University of Ottawa, Ottawa, Ont. K1N 6N5, Canada}

\date{\today}

\begin{abstract}
We demonstrate the controlled creation of a $\mathrm{^{174}Yb^{2+}}$ ion by
photo-ionizing $\mathrm{^{174}Yb^+}$ with weak continuous-wave
lasers at ultraviolet wavelengths.
The photo-ionization is performed by resonantly
exciting transitions of the $\mathrm{^{174}Yb^+}$ ion in three steps.
Starting from an ion crystal of two laser-cooled $\mathrm{^{174}Yb^+}$
ions localized in a radio-frequency trap, the verification of the
ionization process is performed by characterizing the properties
of the resulting mixed-species ion-crystal.
The obtained results facilitate fundamental studies of physics
involving $\mathrm{Yb^{2+}}$ ions.
\end{abstract}


\maketitle
\section{\label{intro}Introduction}

The ionic species $\mathrm{Yb^{2+}}$ has been proposed for several
studies in the field of fundamental physics, namely for tests on the
temporal variation of the fine structure
constant\,\cite{dzuba2003,dzuba2008,Schauer_PRA_82_062518} and for
investigations on the efficient interaction of a single photon with a
single atom in free-space\,\cite{Sondermann_ApplPhysB_89_489}.
For the latter application the isotope $\mathrm{^{174}Yb^{2+}}$ is of
particular relevance, since its level-scheme comprises a nearly
isolated resonance transition which is close to a pure two-level
system.
Trapping such an ion in the focus of a
parabolic mirror\,\cite{maiwald2012collecting} and exciting it with
linear-dipole radiation is expected to enable interaction of light and
matter with almost unit coupling
efficiency~\cite{Sondermann_ApplPhysB_89_489,golla2012,Fischer_APB}.

In this contribution we present the creation of
$\mathrm{^{174}Yb^{2+}}$ starting from a trapped
$\mathrm{^{174}Yb^{+}}$ ion.
The choice of an unfavourable ionization method might
complicate or even hinder the desired experiments due to the creation
of charges in the trapping environment.
In particular, due to the small, sub-wavelength extent of the focal spot of a 
deep parabolic mirror, any charge potential created during the ionization 
process will shift the ion out of the focus of the parabola.  
Therefore, ionization by electron-beam bombardment as performed
in Ref.\,\cite{Schauer_PRA_82_062518} might not be an option.

Previously, several production methods have been applied for the generation of
multiply charged ions.
Most widely used is the production of ions by laser ablation from
solid state
targets\,\cite{Campbell_PRL_102_233004,zhang2001YbIIILifeTime}. 
An alternative is the production of ion-plasmas from vacuum arc
sources\,\cite{Gruber.PhysRevLett.86.636}, potentially followed by
further beam ionization to higher charged states. 
Ions generated with these methods have also been trapped subsequently
to their generation, e.g. in a Penning trap
\cite{Gruber.PhysRevLett.86.636} or in a linear Paul trap
\cite{Campbell_PRL_102_233004}. 
Methods to generate multiply charged ions with a higher level of
control have been demonstrated in
Refs. \cite{Schauer_PRA_82_062518,Feldker_APB_114_11,Kwapien_PRA_75_063418}. 
There, singly charged ions have been generated and trapped inside a
radio-frequency ion-trap. 
The subsequent ionization to the doubly charged state has been
performed by electron-beam bombardment \cite{Schauer_PRA_82_062518},
field ionization using femtosecond laser pulses
\cite{Kwapien_PRA_75_063418}, or non-resonant direct photo-ionization
with a vacuum-ultraviolet light source\,\cite{Feldker_APB_114_11}. 
The work in Ref.\,\cite{Schauer_PRA_82_062518} has to be highlighted
here as the first report on the successful production and trapping of
$\mathrm{Yb^{2+}}$. 
However, all of the aforementioned methods suffer from drawbacks with
respect to our particular experimental scenario.
As already indicated above, methods inducing stray charges are not
acceptable for experiments where light is to be focused tightly onto a
quantum object in a well defined manner.
Furthermore, the weak focusing of all auxiliary beams used in our
set-up (cf. Ref.\,\cite{maiwald2012collecting}) would require
high absolute powers when, e.g. using ionization techniques based on
pulsed lasers.

In contrast, here we demonstrate the resonant photo-ionization of a
single trapped and laser cooled $\mathrm{Yb^{+}}$ ion to
$\mathrm{Yb^{2+}}$, requiring only about $100\,\mathrm{\mu W}$ of
laser power while simultaneously minimizing charging of the trap
environment and obtaining high ionization efficiency. 
Section\,\ref{Sec::Scheme} introduces the pursued ionization scheme
and gives an estimate for the expected ionization rate. 
In Sec.\,\ref{Sec::Experiment} we report the experimental results,
followed by a discussion and an outlook in the last section.

\section{\label{Sec::Scheme}Resonant photo-ionization scheme}
 
\begin{figure}
\centerline{
\includegraphics[scale=0.8]{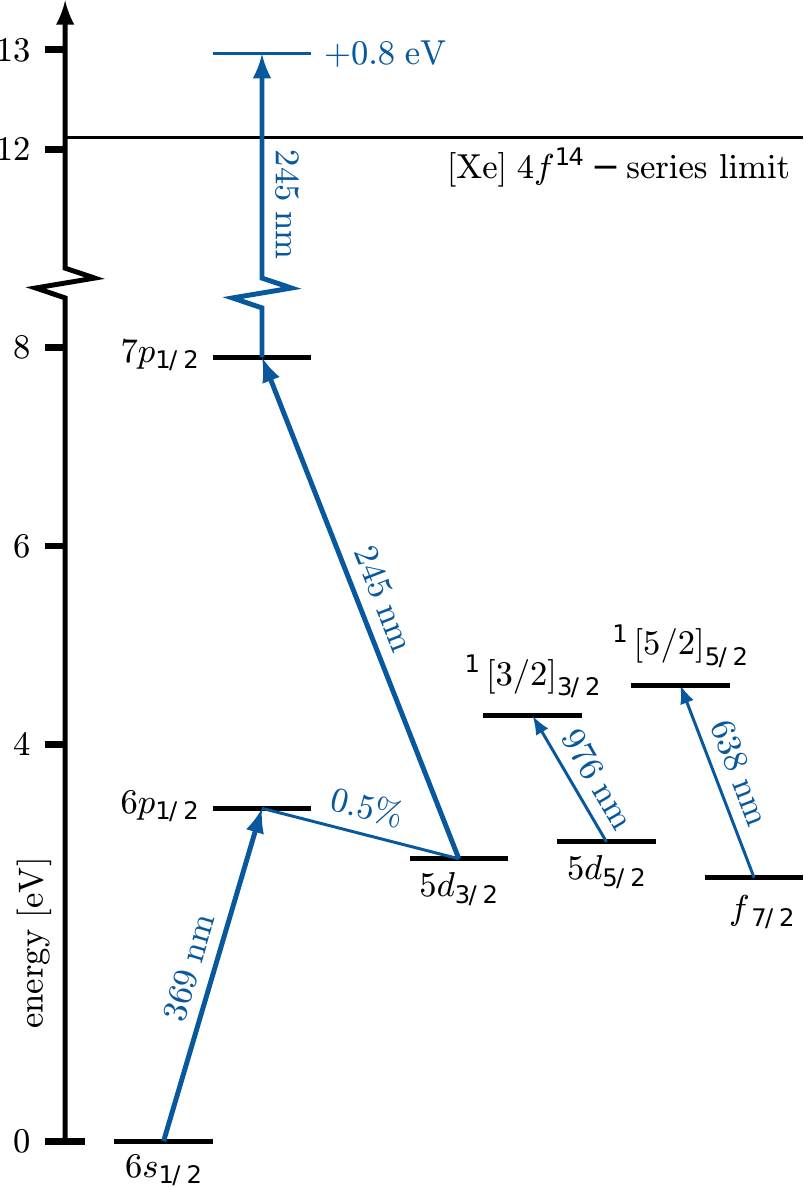}}
\caption{\label{Fig::Scheme}
Level-scheme in $\mathrm{{}^{174}Yb^+}$ relevant for the
photo-ionization process discussed here. 
The $5d_{3/2}$ level is pumped via the $6p_{1/2}$ level using the
transition from the ground state $6s_{1/2}$. 
Another transition from the $5d_{3/2}$ level excites the $7p_{1/2}$ level.
Starting from this $7p_{1/2}$ level a transition to an unbound state
of the single valence electron can be excited using another photon at
$245\NanoMeter$ thereby producing a doubly charged ion.} 
\end{figure}

The photo-ionization scheme used here involves a set of three
transitions as shown in Fig.~\ref{Fig::Scheme}.
In the first step the [Xe]$4f^{14}5d$, J=3/2 level (abbreviated as
$5d_{3/2}$ in the following) 
 is excited via the $6s_{1/2}\rightarrow6p_{1/2}$ transition and a
 subsequent decay $6p_{1/2}\rightarrow5d_{3/2}$.
The latter occurs with a probability
 of $0.5\%$\,\cite{Olmschenk_PhysRevA_76_052314}.
The notation $6s_{1/2}$ is used as an abbreviation for the
  [Xe]$4f^{14}6s$, J=1/2 level and $6p_{1/2}$ for [Xe]$4f^{14}6p$, J=1/2.
In the next step the [Xe]$4f^{14}7p$, J=1/2 level (shortly $7p_{1/2}$)
is excited via the transition starting from $5d_{3/2}$ using light
with a wavelength of $245\NanoMeter$. 
Starting from the $7p_{1/2}$  level a transition towards an unbound
state of the valence electron is realized by excitation with the same
wavelength. 

On the basis of the theoretical data in Ref.\,\cite{DreamPaper} the
decay paths of the $7p_{1/2}$ level can be analyzed. 
The strongest decay with a probability of $69.1\%$ occurs via the
$7p_{1/2}\rightarrow{4f^{14}7s}$, J=1/2 transition from where finally
  either the cooling cycle (see below) or the [Xe]$4f^{14}5d$, J=5/2
  level (shortly $5d_{5/2}$) is entered. 
This level is entered mainly via the intermediate
  [Xe]$4f^{14}6p$, J=3/2 level.
The total probability to enter the $5d_{5/2}$ level via the decay cascade
from the $7p_{1/2}$ level is estimated to be about $5\cdot10^{-3}$.
The decay $7p_{1/2}\rightarrow6s_{1/2}$ occurs with a probability
of $17.7\%$. 
The probability for the decay $7p_{1/2}\rightarrow5d_{3/2}$ is
found to be $12.4\%$. 

The $5d_{5/2}$ level is reported to have a lifetime of
$7.2\MilliSeconds$\,\cite{TaylorInvestigationDtoSClockTransition}. 
Due to this relatively long lifetime the efficiency of the excitation
of the $7p_{1/2}$ level in the above described scheme would be lowered
significantly. 
The $5d_{5/2}$ level is therefore cleared directly via the
$5d_{5/2}\rightarrow{^1[3/2]_{3/2}}$ transition with light at a
wavelength of $976\NanoMeter$.
Here, $^1[3/2]_{3/2}$ is the abbreviation for the
[Xe]$4f^{13}5d6s$, J=3/2 level.   
In addition the $5d_{5/2}$ level can decay to the
[Xe]$4f^{13}6s^2$, J=7/2 level (shortly ${f_{7/2}}$)
with a probability of
$83\%$\,\cite{TaylorInvestigationDtoSClockTransition}.
A lifetime of $3700\PhysicalUnit{d}$ is reported for the latter
level\,\cite{RobertsObservationOfOctupoleTransition}, which is
cleared via the ${f_{7/2}}\rightarrow{^1[5/2]_{5/2}}$ transition
with the wavelength $638\NanoMeter$\,\cite{Gill_PRA_52_R909}, 
where $^1[5/2]_{5/2}$ designates the [Xe]$4f^{13}5d6s$, J=5/2
level.
The upper levels of the transitions driven at $638\NanoMeter$ and
$976\NanoMeter$ decay back into ground state\,\cite{Gill_PRA_52_R909,DreamPaper}.

In the experiment the four transitions mentioned above are driven by
diode-laser based light-sources. 
The inclusion of the two repumping-lasers for the $5d_{5/2}$
and $f_{7/2}$ levels allows for the efficient excitation of the
$7p_{1/2}$ level which in turn maximizes the photoionization probability. 

The performance of this photoionization scheme can be estimated from
the steady-state of the atomic multi-level system excitation
according to the action of the lasers driving the transitions shown in
Fig.~\ref{Fig::Scheme} and the cross-section of the photoionization
from $7p_{1/2}$. 
The rate $R$ for the ionization process can be calculated as:
\begin{equation}
\label{eq-TotalIonizationRate}
R=p_{7p}\sigma_{245}F_{245}\, ,
\end{equation}
with the excitation probability $p_{7p}$ of the $7p_{1/2}$ level, the
photoionization cross-section $\sigma_{245}$ for the photoionization
starting from that level and the flux $F_{245}$ of photons at wavelength
$245\NanoMeter$, which is used for ionization. 

The expected excitation probability  for the $7p_{1/2}$ level has been
computed using the theoretical transition-rate data found in
Ref.\,\cite{DREAM}.  
It was found to be $p_{7p,max}=9.5\cdot10^{-3}$
if all driven transitions are strongly saturated.

The cross-section for the photoionization from $7p_{1/2}$ level has
been calculated following a quantum-defect-theory based
approach\,\cite{Seaton_MonNotRoyAstrSoc_118_504}.  
Refs.\,\cite{Burgess_RevModPhys_30_992,Peach_MemRoyAstrSoc_71_13}
provide two different parametrizations of the corresponding transition
matrix elements. 
The quantum defect values are derived on the basis of the level
energies listed in Ref.\,\cite{NistAtomicSpectraDatabase} and the
[Xe]$4f^{14}$-series ionization-limit from
Ref.\,\cite{Huang_JOSAB_12_961}. 
For the scheme proposed here, these calculations give
$\sigma_{245}=5.5\MegaBarn$ according to 
Ref.\,\cite{Burgess_RevModPhys_30_992} and $\sigma_{245}=7.2\MegaBarn$
according to Ref.\,\cite{Peach_MemRoyAstrSoc_71_13}. 

If the $5d_{3/2}\rightarrow7p_{1/2}$ transition is efficiently
saturated the total ionization rate can be approximated according
  to Eq.\,\ref{eq-TotalIonizationRate} as
\begin{equation}
\label{eq::rate}
R\approx 4.1\cdot10^{-6} \frac{\mathrm{m}^2}{\mathrm{J}} \cdot
\frac{P_{245}}{w_0^2}\, , 
\end{equation}
with a total continuous-wave power of $P_{245}$ for the excitation
laser at wavelength $245\NanoMeter$ and a gaussian beam-waist $w_0$. 
For $w_0=10\MicroMeter$ and $P_{245}=100\MicroWatt$ one finds a total
ionization rate of $R\approx4.1\Seconds^{-1}$. 

We end this section in stressing the importance of the
repumping lasers at $976\NanoMeter$ and $638\NanoMeter$ wavelength.
Starting from the $7p_{1/2}$ level, the ionization to
$\mathrm{Yb^{2+}}$ is calculated to be equally probable as decaying to
the $f_{7/2}$ level for a laser power at $245\NanoMeter$ that is more
than three orders of magnitude larger than the one used in our
experiments. 
For high ionization success probabilities, one would have to increase
the incident power by at least another two orders of magnitude.
As discussed below, at such high power levels detrimental effects
occurred in our experiments.
However, for the low $245\NanoMeter$ laser power leading to successful
ionization the probability for ionization is
considerably lower than for decaying to the branch ending up in the
$f_{7/2}$ level.
Therefore, the laser at $638\NanoMeter$ wavelength clearing this level
is necessary for achieving practical ionization rates in the
experiment.
The same holds true for the $976\NanoMeter$ laser depopulating
the $5d_{5/2}$ level which is the main path for entering the 
$f_{7/2}$ level. 

\section{\label{Sec::Experiment}Experimental results}

\begin{figure}
\centerline{
\includegraphics[width=8cm]{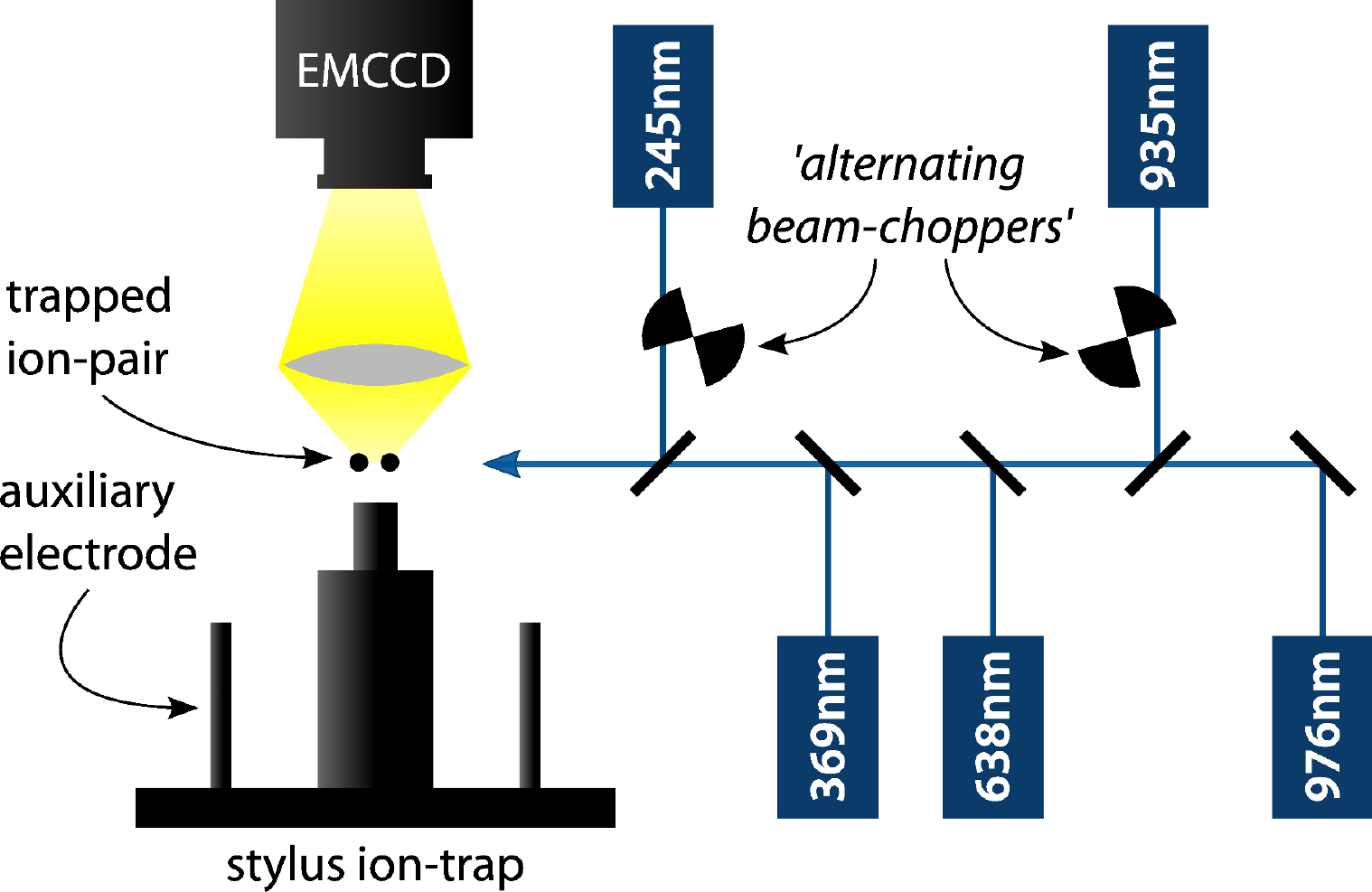}}
\caption{\label{Fig::Setup}
Scheme of the ion-trap setup and laser-configuration for the
ionization experiments. 
All lasers except the laser at the wavelength $245\NanoMeter$ are
applied at a strongly saturated power-level. 
The lasers at $245\NanoMeter$ and $935\NanoMeter$ are alternated using
acousto-optic modulators acting as beam choppers.
The ion motion is probed utilizing one of the auxiliary electrodes.}
\end{figure}

A scheme of the experimental setup for the photoionization-experiments
is shown in Fig.~\ref{Fig::Setup}. 
The ions are confined within a stylus
ion-trap\,\cite{maiwald2009stylus} and the necessary laser-beams are
superimposed and focused onto the ions. 
A set of auxiliary electrodes is employed to compensate stray
electric DC-fields and to probe the motional dynamics of the ion
crystals. For the latter purpose low-amplitude AC signals are
supplied to one of these electrodes.
All involved light-sources are commercially available diode-based
laser-systems based on external-cavity diode lasers tuned by
frequency selective optical feedback.
In addition to the lasers needed for the ionization process,
radiation at $935\NanoMeter$ is needed for laser cooling
Yb$^+$\,\cite{Bell_PRA_44_R20}, where the main cooling
transition is the one at $369\NanoMeter$.
Since light at $935\NanoMeter$ wavelength empties the $5d_{3/2}$ level its
application is detrimental for the ionization process.
Therefore, the $935\NanoMeter$ and $245\NanoMeter$ light paths are
equipped with acousto-optic modulators serving as beam choppers.
This facilitates switching between ionization and cooling cycles.
The fluorescence light at wavelength $369\NanoMeter$ is filtered and
imaged onto a single-photon sensitive camera (EMCCD).
The Yb$^+$ ions themselves are produced by isotope selective
photo-ionization from a stream of neutral Yb atoms. The latter are evaporated from
a resistively heated oven containing Yb foil. The photo-ionization
scheme from neutral Yb to Yb$^+$ is explained in
Refs.\,\cite{hosaka2005,balzer2006}.  

As a first step we investigated the
$5d_{3/2}\rightarrow7p_{1/2}$ transition by applying the laser at
$245\NanoMeter$ wavelength and monitoring the fluorescence from the
$6s_{1/2}\rightarrow6p_{1/2}$ cooling transition.
For this purpose all involved laser powers were chosen such that
an effective saturation parameter $S_{245}\approx0.02$ was reached
when compared to the signal levels obtained from the theoretical
model described above.
The laser at $935\NanoMeter$ is switched off in these experiments,
while the $245\NanoMeter$ beam now serves as a repumper closing the
cooling cycle.

\begin{figure}
\centerline{
  \includegraphics[scale=0.8]{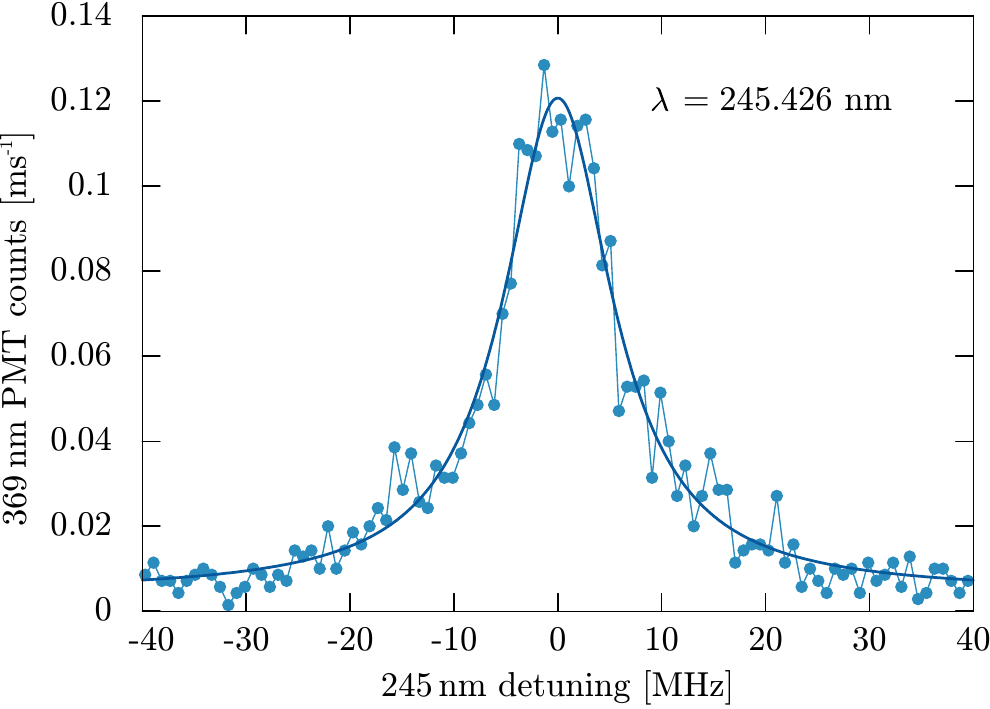}}
  \caption{Scan of the $5d_{3/2}\rightarrow7p_{1/2}$ transition at
    $245\NanoMeter$ with a saturation parameter of
    $S_{245}\approx0.02$. 
  The fluorescence from the decay $6p_{1/2}\rightarrow6s_{1/2}$ at
  $369\NanoMeter$ is detected. 
  The repumping lasers at $638\NanoMeter$ and $976\NanoMeter$ are
  each applied at a strongly saturating power-level. 
  \label{Plot::AomScan245}}
\end{figure}

The result of a spectral scan of the $245\NanoMeter$ laser across 
the $5d_{3/2}\rightarrow7p_{1/2}$ transition is displayed in
Fig.~\ref{Plot::AomScan245}. 
From the fluorescence counts as a function of the laser frequency we
extract a center wavelength of $245.426\NanoMeter$. 
In all subsequent ionization experiments the ionization laser was
tuned to this line center.
From the spectral width of the fluorescence curve a value of
$\tau=13.5\pm2.1\NanoSeconds$ is found for the lifetime of
the $7p_{1/2}$ level.  
No other experimentally obtained values for this lifetime and
wavelength could be found. 
The theoretical data in Ref.\,\cite{DREAM} gives $\tau=24.6\NanoSeconds$,
whereas a value of $\tau=10\NanoSeconds$ can be derived from the data
in Ref.\,\cite{Fawcett_AtDatNucDatTab_47_241}.

The successful ionization in the experimental runs is verified
utilizing a crystallized pair of ions, similar to what has been done
in Refs.~\cite{Kwapien_PRA_75_063418,Feldker_APB_114_11}. 
Initially a pair of $\mathrm{^{174}Yb^+}$ ions is loaded and
laser-cooled into a crystallized state.  
 Following to this step the photoionization
process is started by applying the laser-radiation according to the
scheme shown in Fig.~\ref{Fig::Scheme}. 
During this period the laser at $935\NanoMeter$ is switched off.
Afterwards the outcome of this photoionization attempt is probed using
the laser-cooling scheme with the repumping-laser at $935\NanoMeter$
switched on and the laser at $245\NanoMeter$ switched off. 
In the experiment this sequence has been automated using
accousto-optical modulators with a switching rate of typically
$50\PhysicalUnit{Hz}$. 

\begin{figure}
\centerline{
  \includegraphics[scale=0.8]{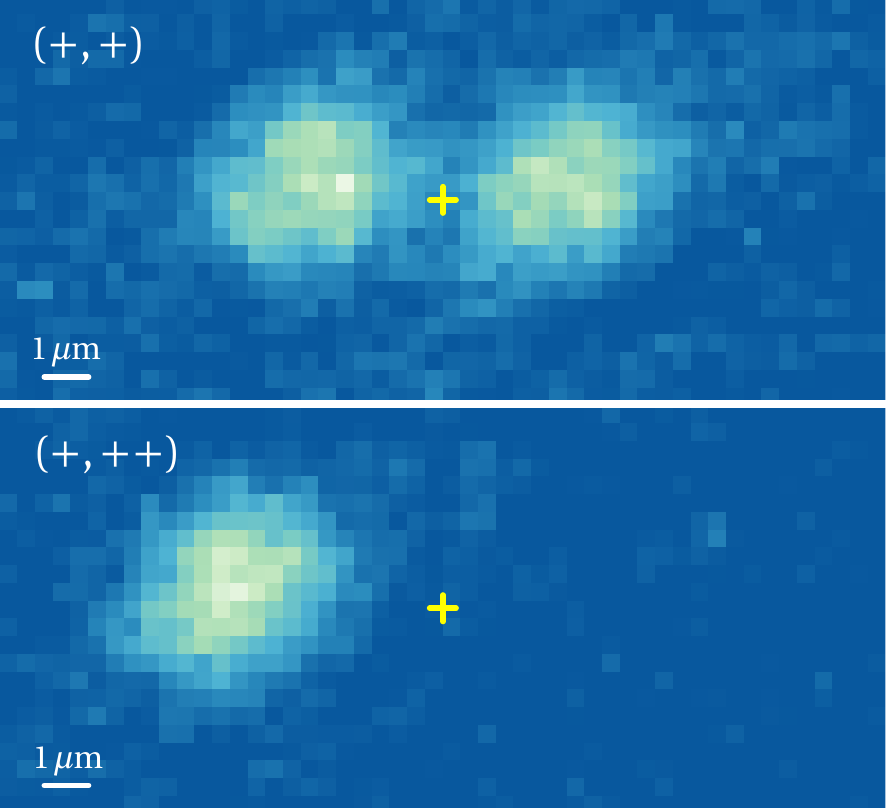}}
  \caption{\label{Image::Ionization}
Images showing the two-ion crystal before and after the application of
the ionization-lasers. 
The yellow cross marks the approximate trap centre.
The increase of the displacement of the remaining singly charged ion
indicates for the increase of charge on the other side of the
crystal. 
}
\end{figure}

Once one of the two ions is turned dark the photoionization sequence
is interrupted and the remaining crystal consisting of a \lq
dark\rq\ and a \lq bright\rq\ ion is analyzed. 
At this point the dark ion remains sympathetically cooled by the
laser cooled bright ion.
A first indication for the successful ionization of the dark ion can
be found in the bright ion's position shift, see also
Refs.~\cite{Kwapien_PRA_75_063418,Feldker_APB_114_11}.
A doubly charged dark ion will be trapped closer to the trap-center
pushing the singly charged bright ion further off. 
This situation is shown in Fig.~\ref{Image::Ionization}. 

The extent of this displacement can be calculated for an arbitrary
trapping potential. 
A two-ion crystal aligns along the weakest axis of the radio-frequency
(RF) ion-trap secular potential. 
The one-dimensional model potential for this situation is given by
\begin{equation}
\label{eq::TwoIonPotential}
V=\sum_{l=1,2}\tfrac{1}{2}M(2\pi\nu_l)^2\, x_l^2(t)+\frac{Q}{\left|
  x_1(t)-x_2(t)\right|}, 
\end{equation}
with the ion mass $M$, the secular-motion frequencies $\nu_l$ 
of ion\,$l$, the factor $Q=q_1q_2/4\pi\varepsilon_0$ and the charges
$q_{1,2}$ of the ions. 
In the following the ion with index $1$ will be assumed to be the
singly charged bright ion. 

We define
\begin{equation}
\eta=\frac{\nu_2}{\nu_1}.
\end{equation}
A value of $\eta=2$ would be expected 
for a singly and a doubly charged ion
if the trapping potential is only constituted of the RF
field. 
This changes in the presence of static electric fields.
Then, due to the difference in the scaling of the resulting forces onto
a particular ion charge the value of $\eta$ will differ from $2$. 

\begin{figure}
\centerline{
  \includegraphics[scale=1.]{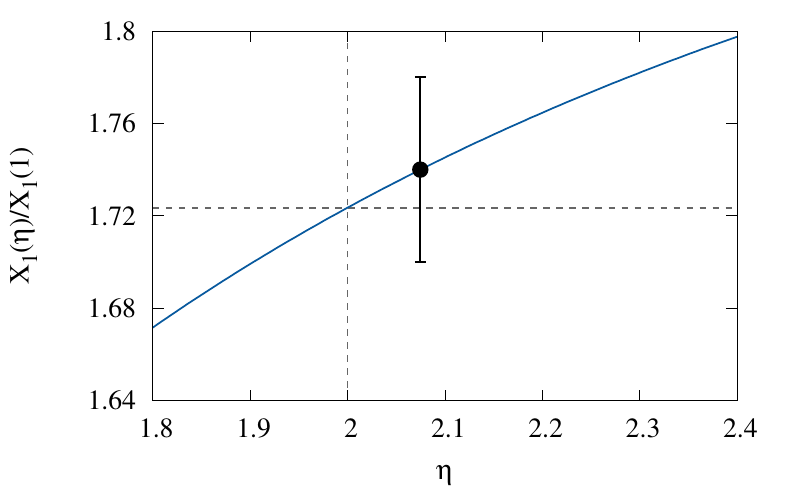}}
  \caption{ \label{Plot::CrystalDistance}
Plot of the normalized displacement $X_1(\eta)/X_1(1)$ of the singly
charged ion from the trap centre as a function of $\eta$,
  assuming $q_1=e$ and $q_2=2e$. 
The symbol marks the value for the displacement ratio of $1.74\pm
  0.04$ as found in the experiment, which would correspond to
  $\eta=2.07$. 
  The dashed lines indicate the ratio expected for $\eta=2$.}
\end{figure}

The equilibrium positions $X_1,X_2$ for each of the two ions are found to be:
\begin{align}
\label{eq::X1}
{X_1}^3&=\frac{Q}{\left(1+\eta^{-2}\right)^{2}M\omega_1^2}\\
X_2&=-\eta^{-2}X_1
\end{align}
with $\omega_1=2\pi\nu_1$.
The plot in Fig.~\ref{Plot::CrystalDistance} shows the normalized
position of the remaining bright ion as a function of $\eta$. 

In the first ionization attempts the laser at $245\NanoMeter$
  was applied at power levels much higher than $100\MicroWatt$.
As suggested by Eq.\,\ref{eq-TotalIonizationRate} and also by
intuition, larger powers should result in larger ionization rates.
Contrary to these expectations we could not observe successful ionization
but a spatial shift of the ions or even ion loss instead.
These effects can be attributed to strong charging of non-conducting
materials in the vacuum chamber induced by the deep-ultraviolet
laser.
For an investigation of charging effects on trapping
performance we refer to Ref.\,\cite{Harlander_NJP_12_093035}.
Hence, the experiments were continued with the ionization laser
operating at low power levels.

Fluorescence images acquired before and after a presumingly successful
ionization attempt are displayed in Fig.~\ref{Image::Ionization}. 
By averaging over a series of such images we obtain the ions'
positions as well as the centre of the trapping potential. 
From these data we deduce a normalized displacement of
$X_1(\eta)/X_1(1)=1.74\pm0.04$, where $X_1(\eta)$ and $X_1(1)$ are the
distances of the bright  $\mathrm{Yb^+}$-ion to the trap centre after
and before the ionization attempt, respectively. 
Within the error margins this displacement is compatible with the
expectation for a $\mathrm{Yb^+}$--$\mathrm{Yb^{2+}}$ crystal, see
Fig.\,\ref{Plot::CrystalDistance}.
A first estimate for the parameter $\eta$ based on this result yields
$\eta=2.07$, while even larger deviations from $\eta=2$ are possible
within the error margins. 
As mentioned above this deviation from $\eta=2$ hints towards the
influence of static electric fields onto the trapping potential.

\begin{figure}
\centerline{
  \includegraphics[scale=0.8]{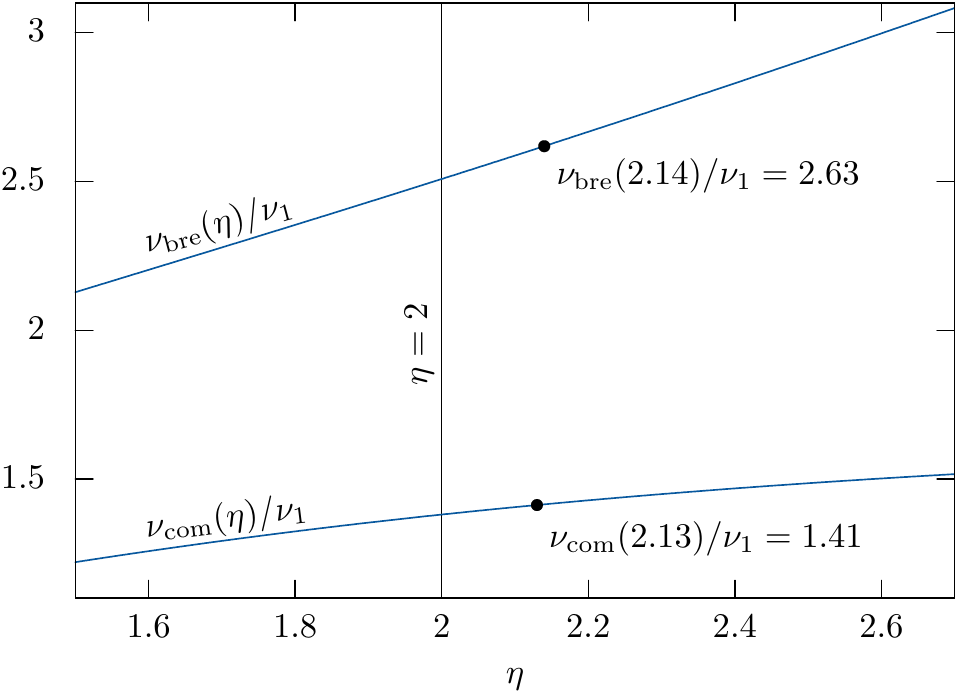}}
  \caption{Plot of the normalized crystal motion frequencies
    $\nu_{\mathrm{bre}}(\eta)/\nu_1$ and
    $\nu_{\mathrm{com}}(\eta)/\nu_1$ as a function of the ratio
    $\eta=\nu_2/\nu_1$ of the single-ion motional-frequencies. 
  The values for $\eta=2.13$ and $\eta=2.14$ as found in the experiment are marked.
  \label{Plot::CrystalFrequencies}}
\end{figure}

The results found above are limited in accuracy by the
discretization of the camera images.
For evaluating $\eta$ more accurately and validating the ionization
success the motional eigenfrequencies of the crystal are investigated,
as has been also done in
Refs.\,\cite{Schauer_PRA_82_062518,Feldker_APB_114_11}.    
For the potential in Eq.\,\ref{eq::TwoIonPotential} the two
crystal-eigenfrequencies can be found to be 
\begin{align}
\nu_{\mathrm{com}}&=\nu_1\cdot
\frac{\eta^4+6 \eta^2-\sqrt{\eta^8+14 \eta^4+1}+1}{2 \eta^2+2}\\
\nu_{\mathrm{br}e}&=\nu_1\cdot
\frac{\eta^4+6 \eta^2+\sqrt{\eta^8+14 \eta^4+1}+1}{2 \eta^2+2}.
\end{align}
The \lq com\rq-mode represents the eigenmode where both ions oscillate in
phase, whereas in the \lq breathe\rq-mode both ions oscillate exactly out of
phase. 
The plots in Fig.~\ref{Plot::CrystalFrequencies} show the dependence
of these frequencies on the parameter $\eta$. 

\begin{figure}
\centerline{
\includegraphics[scale=0.8]{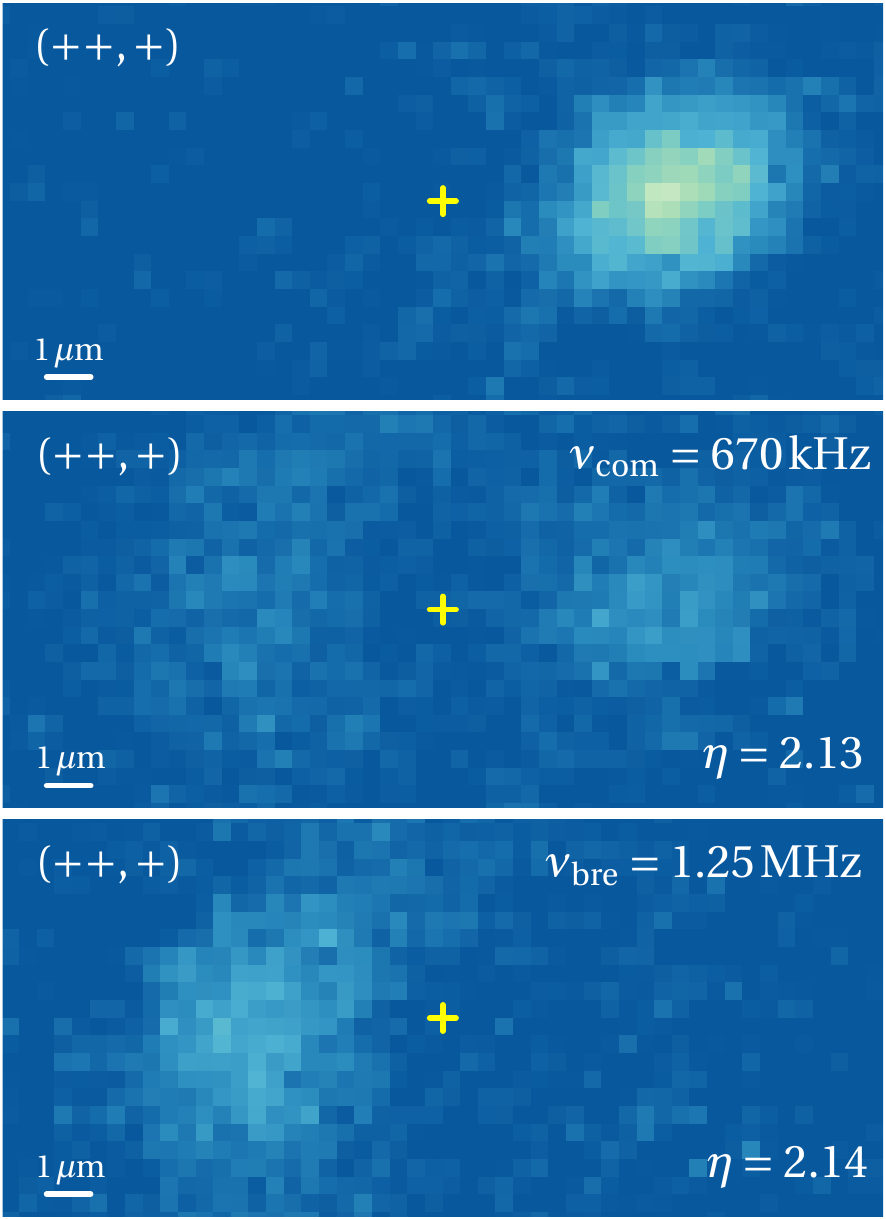}}
\caption{\label{Image::MotionSD}
Series of images showing the excitation of the normal motional
eigenmodes of the mixed-species ion-crystal. 
For the singly charged ion $\nu_1=474\;\mathrm{kHz}$ has been measured.
According to the theoretical model described in the text this gives
$\eta=2.13$ for the common mode and $\eta=2.14$ for the breathing
mode.} 
\end{figure}

In the experiment these crystal-eigenfrequencies are probed by
applying a corresponding RF-electric field using auxilliary
electrodes\,\cite{naegerl1998coherent}. 
Fig.~\ref{Image::MotionSD} shows images of the ions'
  fluorescence during excitation of the two motional
modes along the weak trap axis. 
Initially a value of
$\nu_{\mathrm{com}}(\eta=1)=\nu_1=474\PhysicalUnit{kHz}$ had 
been found for the crystal with two singly charged ions. 
The excitation of the \lq com\rq-mode of the mixed-species crystal gives
$\eta=2.13$. 
From the excitation of the same crystal's \lq breathe\rq-mode $\eta=2.14$ is
found.  
The appearance of $\eta\approx2$ for both modes already hints towards
a successful photoionization of one of the two Yb$^+$ ions. 

As a last step, we combine the results of the measurements on the
spatial shift of the remaining bright ion and the motional
frequencies.
Using Eq.\,\ref{eq::X1}, we determine $q_2$ for the
dark ion from $(X_1(\eta)/X_1(1))^3$ and the average value of
$\eta=2.135$ as found from the motional dynamics of the ion crystal.
This procedure yields $q_2=(1.96\pm 0.14)e$ in good
agreement with a doubly charged ion.

\section{\label{Sec::Discussion} Discussion and Outlook}

The above analysis clarifies that the remaining dark ion is indeed
a doubly charged ytterbium ion.
Even the quantitative agreement of $\eta$ for both
excited crystal modes alone leads to the same conclusion, since
obviously other combinations of $\{\nu_l,\nu_{l^\prime}\}$ can be
ruled out by the observation of the fluorescence from the remaining
Yb$^+$ ion.
The slight deviation from $\eta=2$ can be attributed to the influence
of static electric fields present in the experiment. 
This kind of effect is expected to occur in a stylus-type ion-trap
system where the trapped ions are widely exposed to their
environment. 

In the experiment as described here the photoionization process itself
takes approximately one second. 
Using a laser power of approximately $100\MicroWatt$ for the
ionization laser the process could be run with near
unit efficiency.
For this laser power, the time span of one second until successful
ionization as stated   above constitutes an upper limit, since the
duty cycle for employing the ionization lasers is 50\%. Therefore
the ionization rate is on the order of the one expected from
Eq.~\ref{eq::rate}. 
At larger powers two effects could be observed:
The trapping of the ions is destabilized by the disturbance of
the electric trapping potential from strong photo-electric charges
and, occasionally, dark states of singly charged ytterbium are
produced. 
The mechanism for the production of these dark states could not be
identified yet.

Since the doubly charged species is now readily available,
spectroscopy experiments on the Yb$^{2+}$ 
resonance ${^1S_0}\rightarrow{^3P^\mathrm{o}_1}$ transition at a
wavelength of $252\NanoMeter$ can be initiated. 
As discussed in Ref.\,\cite{Schauer_PRA_82_062518} this is complicated
by the branching of the $^3P^\mathrm{o}_1$ level towards possible
dark-states. 
Hence, the next experimental steps include the spectroscopy on
possible repumping transitions, with the most likely candidate being
the ${^3P^\mathrm{o}_2}\rightarrow{^3P^\mathrm{o}_1}$ transition at
$1578\NanoMeter$ wavelength.

Furthermore, measurements of the life-time of the $^3P^\mathrm{o}_1$
state obtained for Yb$^{2+}$ ions in
plasma~\cite{zhang2001YbIIILifeTime} indicate a maximum possible rate
of fluorescence photons of about $2\cdot10^6/\Seconds$ on full
saturation. 
Therefore, trapping Yb$^{2+}$ ions in a parabolic-mirror trap
system providing high photon-collection
efficiency\,\cite{maiwald2012collecting} is expected to support this
task. 

Finally, we aim at using the Yb$^{2+}$ ions in experiments on efficient
light-matter interaction in free space~\cite{Fischer_APB}, with
the ultimate goal of absorbing a single photon with close to unit
efficiency.

\begin{acknowledgments}
G.L. thanks J. Bergquist for stimulating discussions.
M.S. and G.L. are grateful to the \emph{Deutsche Forschungsgemeinschaft} for
financial support.
G.L. acknowledges financial support from the \emph{European Research
  Council} under the Advanced Grant \lq PACART\rq.
\end{acknowledgments}


\begin{thebibliography}{30}%
\makeatletter
\providecommand \@ifxundefined [1]{%
 \@ifx{#1\undefined}
}%
\providecommand \@ifnum [1]{%
 \ifnum #1\expandafter \@firstoftwo
 \else \expandafter \@secondoftwo
 \fi
}%
\providecommand \@ifx [1]{%
 \ifx #1\expandafter \@firstoftwo
 \else \expandafter \@secondoftwo
 \fi
}%
\providecommand \natexlab [1]{#1}%
\providecommand \enquote  [1]{``#1''}%
\providecommand \bibnamefont  [1]{#1}%
\providecommand \bibfnamefont [1]{#1}%
\providecommand \citenamefont [1]{#1}%
\providecommand \href@noop [0]{\@secondoftwo}%
\providecommand \href [0]{\begingroup \@sanitize@url \@href}%
\providecommand \@href[1]{\@@startlink{#1}\@@href}%
\providecommand \@@href[1]{\endgroup#1\@@endlink}%
\providecommand \@sanitize@url [0]{\catcode `\\12\catcode `\$12\catcode
  `\&12\catcode `\#12\catcode `\^12\catcode `\_12\catcode `\%12\relax}%
\providecommand \@@startlink[1]{}%
\providecommand \@@endlink[0]{}%
\providecommand \url  [0]{\begingroup\@sanitize@url \@url }%
\providecommand \@url [1]{\endgroup\@href {#1}{\urlprefix }}%
\providecommand \urlprefix  [0]{URL }%
\providecommand \Eprint [0]{\href }%
\providecommand \doibase [0]{http://dx.doi.org/}%
\providecommand \selectlanguage [0]{\@gobble}%
\providecommand \bibinfo  [0]{\@secondoftwo}%
\providecommand \bibfield  [0]{\@secondoftwo}%
\providecommand \translation [1]{[#1]}%
\providecommand \BibitemOpen [0]{}%
\providecommand \bibitemStop [0]{}%
\providecommand \bibitemNoStop [0]{.\EOS\space}%
\providecommand \EOS [0]{\spacefactor3000\relax}%
\providecommand \BibitemShut  [1]{\csname bibitem#1\endcsname}%
\let\auto@bib@innerbib\@empty
\bibitem [{\citenamefont {Dzuba}\ \emph {et~al.}(2003)\citenamefont {Dzuba},
  \citenamefont {Flambaum},\ and\ \citenamefont {Marchenko}}]{dzuba2003}%
  \BibitemOpen
  \bibfield  {author} {\bibinfo {author} {\bibfnamefont {V.~A.}\ \bibnamefont
  {Dzuba}}, \bibinfo {author} {\bibfnamefont {V.~V.}\ \bibnamefont {Flambaum}},
  \ and\ \bibinfo {author} {\bibfnamefont {M.~V.}\ \bibnamefont {Marchenko}},\
  }\href {\doibase 10.1103/PhysRevA.68.022506} {\bibfield  {journal} {\bibinfo
  {journal} {Phys. Rev. A}\ }\textbf {\bibinfo {volume} {68}},\ \bibinfo
  {pages} {022506} (\bibinfo {year} {2003})}\BibitemShut {NoStop}%
\bibitem [{\citenamefont {Dzuba}\ and\ \citenamefont
  {Flambaum}(2008)}]{dzuba2008}%
  \BibitemOpen
  \bibfield  {author} {\bibinfo {author} {\bibfnamefont {V.~A.}\ \bibnamefont
  {Dzuba}}\ and\ \bibinfo {author} {\bibfnamefont {V.~V.}\ \bibnamefont
  {Flambaum}},\ }\href {\doibase 10.1103/PhysRevA.77.012515} {\bibfield
  {journal} {\bibinfo  {journal} {Phys. Rev. A}\ }\textbf {\bibinfo {volume}
  {77}},\ \bibinfo {pages} {012515} (\bibinfo {year} {2008})}\BibitemShut
  {NoStop}%
\bibitem [{\citenamefont {Schauer}\ \emph {et~al.}(2010)\citenamefont
  {Schauer}, \citenamefont {Danielson}, \citenamefont {Feldbaum}, \citenamefont
  {Rahaman}, \citenamefont {Wang}, \citenamefont {Zhang}, \citenamefont
  {Zhao},\ and\ \citenamefont {Torgerson}}]{Schauer_PRA_82_062518}%
  \BibitemOpen
  \bibfield  {author} {\bibinfo {author} {\bibfnamefont {M.~M.}\ \bibnamefont
  {Schauer}}, \bibinfo {author} {\bibfnamefont {J.~R.}\ \bibnamefont
  {Danielson}}, \bibinfo {author} {\bibfnamefont {D.}~\bibnamefont {Feldbaum}},
  \bibinfo {author} {\bibfnamefont {M.~S.}\ \bibnamefont {Rahaman}}, \bibinfo
  {author} {\bibfnamefont {L.-B.}\ \bibnamefont {Wang}}, \bibinfo {author}
  {\bibfnamefont {J.}~\bibnamefont {Zhang}}, \bibinfo {author} {\bibfnamefont
  {X.}~\bibnamefont {Zhao}}, \ and\ \bibinfo {author} {\bibfnamefont {J.~R.}\
  \bibnamefont {Torgerson}},\ }\href {\doibase 10.1103/PhysRevA.82.062518}
  {\bibfield  {journal} {\bibinfo  {journal} {Phys. Rev. A}\ }\textbf {\bibinfo
  {volume} {82}},\ \bibinfo {pages} {062518} (\bibinfo {year}
  {2010})}\BibitemShut {NoStop}%
\bibitem [{\citenamefont {Sondermann}\ \emph {et~al.}(2007)\citenamefont
  {Sondermann}, \citenamefont {Maiwald}, \citenamefont {Konermann},
  \citenamefont {Lindlein}, \citenamefont {Peschel},\ and\ \citenamefont
  {Leuchs}}]{Sondermann_ApplPhysB_89_489}%
  \BibitemOpen
  \bibfield  {author} {\bibinfo {author} {\bibfnamefont {M.}~\bibnamefont
  {Sondermann}}, \bibinfo {author} {\bibfnamefont {R.}~\bibnamefont {Maiwald}},
  \bibinfo {author} {\bibfnamefont {H.}~\bibnamefont {Konermann}}, \bibinfo
  {author} {\bibfnamefont {N.}~\bibnamefont {Lindlein}}, \bibinfo {author}
  {\bibfnamefont {U.}~\bibnamefont {Peschel}}, \ and\ \bibinfo {author}
  {\bibfnamefont {G.}~\bibnamefont {Leuchs}},\ }\href@noop {} {\bibfield
  {journal} {\bibinfo  {journal} {Applied Physics B}\ }\textbf {\bibinfo
  {volume} {89}},\ \bibinfo {pages} {489} (\bibinfo {year} {2007})}\BibitemShut
  {NoStop}%
\bibitem [{\citenamefont {Maiwald}\ \emph {et~al.}(2012)\citenamefont
  {Maiwald}, \citenamefont {Golla}, \citenamefont {Fischer}, \citenamefont
  {Bader}, \citenamefont {Heugel}, \citenamefont {Chalopin}, \citenamefont
  {Sondermann},\ and\ \citenamefont {Leuchs}}]{maiwald2012collecting}%
  \BibitemOpen
  \bibfield  {author} {\bibinfo {author} {\bibfnamefont {R.}~\bibnamefont
  {Maiwald}}, \bibinfo {author} {\bibfnamefont {A.}~\bibnamefont {Golla}},
  \bibinfo {author} {\bibfnamefont {M.}~\bibnamefont {Fischer}}, \bibinfo
  {author} {\bibfnamefont {M.}~\bibnamefont {Bader}}, \bibinfo {author}
  {\bibfnamefont {S.}~\bibnamefont {Heugel}}, \bibinfo {author} {\bibfnamefont
  {B.}~\bibnamefont {Chalopin}}, \bibinfo {author} {\bibfnamefont
  {M.}~\bibnamefont {Sondermann}}, \ and\ \bibinfo {author} {\bibfnamefont
  {G.}~\bibnamefont {Leuchs}},\ }\href@noop {} {\bibfield  {journal} {\bibinfo
  {journal} {Physical Review A}\ }\textbf {\bibinfo {volume} {86}},\ \bibinfo
  {pages} {043431} (\bibinfo {year} {2012})}\BibitemShut {NoStop}%
\bibitem [{\citenamefont {Golla}\ \emph {et~al.}(2012)\citenamefont {Golla},
  \citenamefont {Chalopin}, \citenamefont {Bader}, \citenamefont {Harder},
  \citenamefont {Mantel}, \citenamefont {Maiwald}, \citenamefont {Lindlein},
  \citenamefont {Sondermann},\ and\ \citenamefont {Leuchs}}]{golla2012}%
  \BibitemOpen
  \bibfield  {author} {\bibinfo {author} {\bibfnamefont {A.}~\bibnamefont
  {Golla}}, \bibinfo {author} {\bibfnamefont {B.}~\bibnamefont {Chalopin}},
  \bibinfo {author} {\bibfnamefont {M.}~\bibnamefont {Bader}}, \bibinfo
  {author} {\bibfnamefont {I.}~\bibnamefont {Harder}}, \bibinfo {author}
  {\bibfnamefont {K.}~\bibnamefont {Mantel}}, \bibinfo {author} {\bibfnamefont
  {R.}~\bibnamefont {Maiwald}}, \bibinfo {author} {\bibfnamefont
  {N.}~\bibnamefont {Lindlein}}, \bibinfo {author} {\bibfnamefont
  {M.}~\bibnamefont {Sondermann}}, \ and\ \bibinfo {author} {\bibfnamefont
  {G.}~\bibnamefont {Leuchs}},\ }\href {\doibase 10.1140/epjd/e2012-30293-y}
  {\bibfield  {journal} {\bibinfo  {journal} {Eur. Phys. J. D}\ }\textbf
  {\bibinfo {volume} {66}},\ \bibinfo {pages} {190} (\bibinfo {year} {2012})},\
  \Eprint {http://arxiv.org/abs/arXiv:1207.3215} {arXiv:1207.3215} \BibitemShut
  {NoStop}%
\bibitem [{\citenamefont {Fischer}\ \emph {et~al.}(2014)\citenamefont
  {Fischer}, \citenamefont {Bader}, \citenamefont {Maiwald}, \citenamefont
  {Golla}, \citenamefont {Sondermann},\ and\ \citenamefont
  {Leuchs}}]{Fischer_APB}%
  \BibitemOpen
  \bibfield  {author} {\bibinfo {author} {\bibfnamefont {M.}~\bibnamefont
  {Fischer}}, \bibinfo {author} {\bibfnamefont {M.}~\bibnamefont {Bader}},
  \bibinfo {author} {\bibfnamefont {R.}~\bibnamefont {Maiwald}}, \bibinfo
  {author} {\bibfnamefont {A.}~\bibnamefont {Golla}}, \bibinfo {author}
  {\bibfnamefont {M.}~\bibnamefont {Sondermann}}, \ and\ \bibinfo {author}
  {\bibfnamefont {G.}~\bibnamefont {Leuchs}},\ }\href {\doibase
  10.1007/s00340-014-5817-y} {\bibfield  {journal} {\bibinfo  {journal} {Appl.
  Phys. B}\ }\textbf {\bibinfo {volume} {117}},\ \bibinfo {pages} {797}
  (\bibinfo {year} {2014})},\ \Eprint {http://arxiv.org/abs/arXiv:1311.1982}
  {arXiv:1311.1982} \BibitemShut {NoStop}%
\bibitem [{\citenamefont {Campbell}\ \emph {et~al.}(2009)\citenamefont
  {Campbell}, \citenamefont {Steele}, \citenamefont {Churchill}, \citenamefont
  {DePalatis}, \citenamefont {Naylor}, \citenamefont {Matsukevich},
  \citenamefont {Kuzmich},\ and\ \citenamefont
  {Chapman}}]{Campbell_PRL_102_233004}%
  \BibitemOpen
  \bibfield  {author} {\bibinfo {author} {\bibfnamefont {C.}~\bibnamefont
  {Campbell}}, \bibinfo {author} {\bibfnamefont {A.}~\bibnamefont {Steele}},
  \bibinfo {author} {\bibfnamefont {L.}~\bibnamefont {Churchill}}, \bibinfo
  {author} {\bibfnamefont {M.}~\bibnamefont {DePalatis}}, \bibinfo {author}
  {\bibfnamefont {D.}~\bibnamefont {Naylor}}, \bibinfo {author} {\bibfnamefont
  {D.}~\bibnamefont {Matsukevich}}, \bibinfo {author} {\bibfnamefont
  {A.}~\bibnamefont {Kuzmich}}, \ and\ \bibinfo {author} {\bibfnamefont
  {M.}~\bibnamefont {Chapman}},\ }\href@noop {} {\bibfield  {journal} {\bibinfo
   {journal} {Physical Review Letters}\ }\textbf {\bibinfo {volume} {102}},\
  \bibinfo {pages} {233004} (\bibinfo {year} {2009})}\BibitemShut {NoStop}%
\bibitem [{\citenamefont {Zhang}\ \emph {et~al.}(2001)\citenamefont {Zhang},
  \citenamefont {Li}, \citenamefont {Svanberg}, \citenamefont {Palmeri},
  \citenamefont {Quinet},\ and\ \citenamefont
  {Biemont}}]{zhang2001YbIIILifeTime}%
  \BibitemOpen
  \bibfield  {author} {\bibinfo {author} {\bibfnamefont {Z.~G.}\ \bibnamefont
  {Zhang}}, \bibinfo {author} {\bibfnamefont {Z.~S.}\ \bibnamefont {Li}},
  \bibinfo {author} {\bibfnamefont {S.}~\bibnamefont {Svanberg}}, \bibinfo
  {author} {\bibfnamefont {P.}~\bibnamefont {Palmeri}}, \bibinfo {author}
  {\bibfnamefont {P.}~\bibnamefont {Quinet}}, \ and\ \bibinfo {author}
  {\bibfnamefont {E.}~\bibnamefont {Biemont}},\ }\href@noop {} {\bibfield
  {journal} {\bibinfo  {journal} {Eur. Phys. J. D}\ }\textbf {\bibinfo {volume}
  {15}},\ \bibinfo {pages} {301} (\bibinfo {year} {2001})}\BibitemShut
  {NoStop}%
\bibitem [{\citenamefont {Gruber}\ \emph {et~al.}(2001)\citenamefont {Gruber},
  \citenamefont {Holder}, \citenamefont {Steiger}, \citenamefont {Beck},
  \citenamefont {DeWitt}, \citenamefont {Glassman}, \citenamefont {McDonald},
  \citenamefont {Church},\ and\ \citenamefont
  {Schneider}}]{Gruber.PhysRevLett.86.636}%
  \BibitemOpen
  \bibfield  {author} {\bibinfo {author} {\bibfnamefont {L.}~\bibnamefont
  {Gruber}}, \bibinfo {author} {\bibfnamefont {J.}~\bibnamefont {Holder}},
  \bibinfo {author} {\bibfnamefont {J.}~\bibnamefont {Steiger}}, \bibinfo
  {author} {\bibfnamefont {B.}~\bibnamefont {Beck}}, \bibinfo {author}
  {\bibfnamefont {H.}~\bibnamefont {DeWitt}}, \bibinfo {author} {\bibfnamefont
  {J.}~\bibnamefont {Glassman}}, \bibinfo {author} {\bibfnamefont
  {J.}~\bibnamefont {McDonald}}, \bibinfo {author} {\bibfnamefont
  {D.}~\bibnamefont {Church}}, \ and\ \bibinfo {author} {\bibfnamefont
  {D.}~\bibnamefont {Schneider}},\ }\href@noop {} {\bibfield  {journal}
  {\bibinfo  {journal} {Phys. Rev. Lett.}\ }\textbf {\bibinfo {volume} {86}},\
  \bibinfo {pages} {636} (\bibinfo {year} {2001})}\BibitemShut {NoStop}%
\bibitem [{\citenamefont {Feldker}\ \emph {et~al.}(2014)\citenamefont
  {Feldker}, \citenamefont {Pelzer}, \citenamefont {Stappel}, \citenamefont
  {Bachor}, \citenamefont {Kolbe}, \citenamefont {Walz},\ and\ \citenamefont
  {Schmidt-Kaler}}]{Feldker_APB_114_11}%
  \BibitemOpen
  \bibfield  {author} {\bibinfo {author} {\bibfnamefont {T.}~\bibnamefont
  {Feldker}}, \bibinfo {author} {\bibfnamefont {L.}~\bibnamefont {Pelzer}},
  \bibinfo {author} {\bibfnamefont {M.}~\bibnamefont {Stappel}}, \bibinfo
  {author} {\bibfnamefont {P.}~\bibnamefont {Bachor}}, \bibinfo {author}
  {\bibfnamefont {D.}~\bibnamefont {Kolbe}}, \bibinfo {author} {\bibfnamefont
  {J.}~\bibnamefont {Walz}}, \ and\ \bibinfo {author} {\bibfnamefont
  {F.}~\bibnamefont {Schmidt-Kaler}},\ }\href@noop {} {\bibfield  {journal}
  {\bibinfo  {journal} {Applied Physics B}\ }\textbf {\bibinfo {volume}
  {114}},\ \bibinfo {pages} {11} (\bibinfo {year} {2014})}\BibitemShut
  {NoStop}%
\bibitem [{\citenamefont {Kwapie\ifmmode~\acute{n}\else \'{n}\fi{}}\ \emph
  {et~al.}(2007)\citenamefont {Kwapie\ifmmode~\acute{n}\else \'{n}\fi{}},
  \citenamefont {Eichmann},\ and\ \citenamefont
  {Sandner}}]{Kwapien_PRA_75_063418}%
  \BibitemOpen
  \bibfield  {author} {\bibinfo {author} {\bibfnamefont {T.}~\bibnamefont
  {Kwapie\ifmmode~\acute{n}\else \'{n}\fi{}}}, \bibinfo {author} {\bibfnamefont
  {U.}~\bibnamefont {Eichmann}}, \ and\ \bibinfo {author} {\bibfnamefont
  {W.}~\bibnamefont {Sandner}},\ }\href {\doibase 10.1103/PhysRevA.75.063418}
  {\bibfield  {journal} {\bibinfo  {journal} {Phys. Rev. A}\ }\textbf {\bibinfo
  {volume} {75}},\ \bibinfo {pages} {063418} (\bibinfo {year}
  {2007})}\BibitemShut {NoStop}%
\bibitem [{\citenamefont {Olmschenk}\ \emph {et~al.}(2007)\citenamefont
  {Olmschenk}, \citenamefont {Younge}, \citenamefont {Moehring}, \citenamefont
  {Matsukevich}, \citenamefont {Maunz},\ and\ \citenamefont
  {Monroe}}]{Olmschenk_PhysRevA_76_052314}%
  \BibitemOpen
  \bibfield  {author} {\bibinfo {author} {\bibfnamefont {S.}~\bibnamefont
  {Olmschenk}}, \bibinfo {author} {\bibfnamefont {K.~C.}\ \bibnamefont
  {Younge}}, \bibinfo {author} {\bibfnamefont {D.~L.}\ \bibnamefont
  {Moehring}}, \bibinfo {author} {\bibfnamefont {D.~N.}\ \bibnamefont
  {Matsukevich}}, \bibinfo {author} {\bibfnamefont {P.}~\bibnamefont {Maunz}},
  \ and\ \bibinfo {author} {\bibfnamefont {C.}~\bibnamefont {Monroe}},\ }\href
  {\doibase 10.1103/PhysRevA.76.052314} {\bibfield  {journal} {\bibinfo
  {journal} {Phys. Rev. A}\ }\textbf {\bibinfo {volume} {76}},\ \bibinfo
  {pages} {052314} (\bibinfo {year} {2007})}\BibitemShut {NoStop}%
\bibitem [{\citenamefont {Bi\'emont}\ \emph {et~al.}(2000)\citenamefont
  {Bi\'emont}, \citenamefont {Palmeri},\ and\ \citenamefont
  {Quinet}}]{DreamPaper}%
  \BibitemOpen
  \bibfield  {author} {\bibinfo {author} {\bibfnamefont {E.}~\bibnamefont
  {Bi\'emont}}, \bibinfo {author} {\bibfnamefont {P.}~\bibnamefont {Palmeri}},
  \ and\ \bibinfo {author} {\bibfnamefont {P.}~\bibnamefont {Quinet}},\ }in\
  \href {\doibase 10.1007/978-94-011-4114-7_64} {\emph {\bibinfo {booktitle}
  {Toward a New Millennium in Galaxy Morphology}}},\ \bibinfo {editor} {edited
  by\ \bibinfo {editor} {\bibfnamefont {D.}~\bibnamefont {Block}}, \bibinfo
  {editor} {\bibfnamefont {I.}~\bibnamefont {Puerari}}, \bibinfo {editor}
  {\bibfnamefont {A.}~\bibnamefont {Stockton}}, \ and\ \bibinfo {editor}
  {\bibfnamefont {D.}~\bibnamefont {Ferreira}}}\ (\bibinfo  {publisher}
  {Springer Netherlands},\ \bibinfo {year} {2000})\ pp.\ \bibinfo {pages}
  {635--637}\BibitemShut {NoStop}%
\bibitem [{\citenamefont {Taylor}\ \emph {et~al.}(1997)\citenamefont {Taylor},
  \citenamefont {Roberts}, \citenamefont {Gateva-Kostova}, \citenamefont
  {Clarke}, \citenamefont {Barwood}, \citenamefont {Rowley},\ and\
  \citenamefont {Gill}}]{TaylorInvestigationDtoSClockTransition}%
  \BibitemOpen
  \bibfield  {author} {\bibinfo {author} {\bibfnamefont {P.}~\bibnamefont
  {Taylor}}, \bibinfo {author} {\bibfnamefont {M.}~\bibnamefont {Roberts}},
  \bibinfo {author} {\bibfnamefont {S.~V.}\ \bibnamefont {Gateva-Kostova}},
  \bibinfo {author} {\bibfnamefont {R.~B.~M.}\ \bibnamefont {Clarke}}, \bibinfo
  {author} {\bibfnamefont {G.~P.}\ \bibnamefont {Barwood}}, \bibinfo {author}
  {\bibfnamefont {W.~R.~C.}\ \bibnamefont {Rowley}}, \ and\ \bibinfo {author}
  {\bibfnamefont {P.}~\bibnamefont {Gill}},\ }\href@noop {} {\bibfield
  {journal} {\bibinfo  {journal} {Phys. Rev. A}\ }\textbf {\bibinfo {volume}
  {56}},\ \bibinfo {pages} {2699} (\bibinfo {year} {1997})}\BibitemShut
  {NoStop}%
\bibitem [{\citenamefont {Roberts}\ \emph {et~al.}(1997)\citenamefont
  {Roberts}, \citenamefont {Taylor}, \citenamefont {Barwood}, \citenamefont
  {Gill}, \citenamefont {Klein},\ and\ \citenamefont
  {Rowley}}]{RobertsObservationOfOctupoleTransition}%
  \BibitemOpen
  \bibfield  {author} {\bibinfo {author} {\bibfnamefont {M.}~\bibnamefont
  {Roberts}}, \bibinfo {author} {\bibfnamefont {P.}~\bibnamefont {Taylor}},
  \bibinfo {author} {\bibfnamefont {G.~P.}\ \bibnamefont {Barwood}}, \bibinfo
  {author} {\bibfnamefont {P.}~\bibnamefont {Gill}}, \bibinfo {author}
  {\bibfnamefont {H.~A.}\ \bibnamefont {Klein}}, \ and\ \bibinfo {author}
  {\bibfnamefont {W.~R.~C.}\ \bibnamefont {Rowley}},\ }\href@noop {} {\bibfield
   {journal} {\bibinfo  {journal} {Phys. Rev. Lett.}\ }\textbf {\bibinfo
  {volume} {78}},\ \bibinfo {pages} {1876} (\bibinfo {year}
  {1997})}\BibitemShut {NoStop}%
\bibitem [{\citenamefont {Gill}\ \emph {et~al.}(1995)\citenamefont {Gill},
  \citenamefont {Klein}, \citenamefont {Levick}, \citenamefont {Roberts},
  \citenamefont {Rowley},\ and\ \citenamefont {Taylor}}]{Gill_PRA_52_R909}%
  \BibitemOpen
  \bibfield  {author} {\bibinfo {author} {\bibfnamefont {P.}~\bibnamefont
  {Gill}}, \bibinfo {author} {\bibfnamefont {H.~A.}\ \bibnamefont {Klein}},
  \bibinfo {author} {\bibfnamefont {A.~P.}\ \bibnamefont {Levick}}, \bibinfo
  {author} {\bibfnamefont {M.}~\bibnamefont {Roberts}}, \bibinfo {author}
  {\bibfnamefont {W.~R.~C.}\ \bibnamefont {Rowley}}, \ and\ \bibinfo {author}
  {\bibfnamefont {P.}~\bibnamefont {Taylor}},\ }\href {\doibase
  10.1103/PhysRevA.52.R909} {\bibfield  {journal} {\bibinfo  {journal} {Phys.
  Rev. A}\ }\textbf {\bibinfo {volume} {52}},\ \bibinfo {pages} {R909}
  (\bibinfo {year} {1995})}\BibitemShut {NoStop}%
\bibitem [{\citenamefont {Bi\'emont}\ \emph {et~al.}(2012)\citenamefont
  {Bi\'emont}, \citenamefont {Palmeri},\ and\ \citenamefont {Quinet}}]{DREAM}%
  \BibitemOpen
  \bibfield  {author} {\bibinfo {author} {\bibfnamefont {E.}~\bibnamefont
  {Bi\'emont}}, \bibinfo {author} {\bibfnamefont {P.}~\bibnamefont {Palmeri}},
  \ and\ \bibinfo {author} {\bibfnamefont {P.}~\bibnamefont {Quinet}},\
  }\href@noop {} {\enquote {\bibinfo {title} {D.r.e.a.m., database on rare
  earths at mons university, http://w3.umons.ac.be/astro/dream.shtml},}\
  }\bibinfo {howpublished} {\url{http://w3.umons.ac.be/astro/dream.shtml}}
  (\bibinfo {year} {2012})\BibitemShut {NoStop}%
\bibitem [{\citenamefont {Seaton}(1958)}]{Seaton_MonNotRoyAstrSoc_118_504}%
  \BibitemOpen
  \bibfield  {author} {\bibinfo {author} {\bibfnamefont {M.}~\bibnamefont
  {Seaton}},\ }\href@noop {} {\bibfield  {journal} {\bibinfo  {journal}
  {Monthly Notices of the Royal Astronomical Society}\ }\textbf {\bibinfo
  {volume} {118}},\ \bibinfo {pages} {504} (\bibinfo {year}
  {1958})}\BibitemShut {NoStop}%
\bibitem [{\citenamefont {Burgess}\ and\ \citenamefont
  {Seaton}(1958)}]{Burgess_RevModPhys_30_992}%
  \BibitemOpen
  \bibfield  {author} {\bibinfo {author} {\bibfnamefont {A.}~\bibnamefont
  {Burgess}}\ and\ \bibinfo {author} {\bibfnamefont {M.}~\bibnamefont
  {Seaton}},\ }\href@noop {} {\bibfield  {journal} {\bibinfo  {journal}
  {Reviews of Modern Physics}\ }\textbf {\bibinfo {volume} {30}},\ \bibinfo
  {pages} {992} (\bibinfo {year} {1958})}\BibitemShut {NoStop}%
\bibitem [{\citenamefont {Peach}(1967)}]{Peach_MemRoyAstrSoc_71_13}%
  \BibitemOpen
  \bibfield  {author} {\bibinfo {author} {\bibfnamefont {G.}~\bibnamefont
  {Peach}},\ }\href@noop {} {\bibfield  {journal} {\bibinfo  {journal} {Memoirs
  of the Royal Astronomical Society}\ }\textbf {\bibinfo {volume} {71}},\
  \bibinfo {pages} {13} (\bibinfo {year} {1967})}\BibitemShut {NoStop}%
\bibitem [{\citenamefont {Ralchenko}\ \emph {et~al.}(2012)\citenamefont
  {Ralchenko}, \citenamefont {Kramida}, \citenamefont {Reader},\ and\
  \citenamefont {(2011)}}]{NistAtomicSpectraDatabase}%
  \BibitemOpen
  \bibfield  {author} {\bibinfo {author} {\bibfnamefont {Y.}~\bibnamefont
  {Ralchenko}}, \bibinfo {author} {\bibfnamefont {A.}~\bibnamefont {Kramida}},
  \bibinfo {author} {\bibfnamefont {J.}~\bibnamefont {Reader}}, \ and\ \bibinfo
  {author} {\bibfnamefont {N.~A.~T.}\ \bibnamefont {(2011)}},\ }\href@noop {}
  {\enquote {\bibinfo {title} {Nist atomic spectra database (ver. 4.1.0),
  [online], http://physics.nist.gov/asd},}\ }\bibinfo {howpublished}
  {http://physics.nist.gov/asd} (\bibinfo {year} {2012})\BibitemShut {NoStop}%
\bibitem [{\citenamefont {Huang}\ \emph {et~al.}(1995)\citenamefont {Huang},
  \citenamefont {Xu}, \citenamefont {Xu}, \citenamefont {Xue},\ and\
  \citenamefont {Chen}}]{Huang_JOSAB_12_961}%
  \BibitemOpen
  \bibfield  {author} {\bibinfo {author} {\bibfnamefont {W.}~\bibnamefont
  {Huang}}, \bibinfo {author} {\bibfnamefont {X.}~\bibnamefont {Xu}}, \bibinfo
  {author} {\bibfnamefont {C.}~\bibnamefont {Xu}}, \bibinfo {author}
  {\bibfnamefont {M.}~\bibnamefont {Xue}}, \ and\ \bibinfo {author}
  {\bibfnamefont {D.}~\bibnamefont {Chen}},\ }\href@noop {} {\bibfield
  {journal} {\bibinfo  {journal} {J. Opt. Soc. Am. B}\ }\textbf {\bibinfo
  {volume} {12}},\ \bibinfo {pages} {961} (\bibinfo {year} {1995})}\BibitemShut
  {NoStop}%
\bibitem [{\citenamefont {Maiwald}\ \emph {et~al.}(2009)\citenamefont
  {Maiwald}, \citenamefont {Leibfried}, \citenamefont {Britton}, \citenamefont
  {Bergquist}, \citenamefont {Leuchs},\ and\ \citenamefont
  {Wineland}}]{maiwald2009stylus}%
  \BibitemOpen
  \bibfield  {author} {\bibinfo {author} {\bibfnamefont {R.}~\bibnamefont
  {Maiwald}}, \bibinfo {author} {\bibfnamefont {D.}~\bibnamefont {Leibfried}},
  \bibinfo {author} {\bibfnamefont {J.}~\bibnamefont {Britton}}, \bibinfo
  {author} {\bibfnamefont {J.~C.}\ \bibnamefont {Bergquist}}, \bibinfo {author}
  {\bibfnamefont {G.}~\bibnamefont {Leuchs}}, \ and\ \bibinfo {author}
  {\bibfnamefont {D.~J.}\ \bibnamefont {Wineland}},\ }\href@noop {} {\bibfield
  {journal} {\bibinfo  {journal} {Nature Physics}\ }\textbf {\bibinfo {volume}
  {5}},\ \bibinfo {pages} {551} (\bibinfo {year} {2009})}\BibitemShut {NoStop}%
\bibitem [{\citenamefont {Bell}\ \emph {et~al.}(1991)\citenamefont {Bell},
  \citenamefont {Gill}, \citenamefont {Klein}, \citenamefont {Levick},
  \citenamefont {Tamm},\ and\ \citenamefont {Schnier}}]{Bell_PRA_44_R20}%
  \BibitemOpen
  \bibfield  {author} {\bibinfo {author} {\bibfnamefont {A.~S.}\ \bibnamefont
  {Bell}}, \bibinfo {author} {\bibfnamefont {P.}~\bibnamefont {Gill}}, \bibinfo
  {author} {\bibfnamefont {H.~A.}\ \bibnamefont {Klein}}, \bibinfo {author}
  {\bibfnamefont {A.~P.}\ \bibnamefont {Levick}}, \bibinfo {author}
  {\bibfnamefont {C.}~\bibnamefont {Tamm}}, \ and\ \bibinfo {author}
  {\bibfnamefont {D.}~\bibnamefont {Schnier}},\ }\href {\doibase
  10.1103/PhysRevA.44.R20} {\bibfield  {journal} {\bibinfo  {journal} {Phys.
  Rev. A}\ }\textbf {\bibinfo {volume} {44}},\ \bibinfo {pages} {R20} (\bibinfo
  {year} {1991})}\BibitemShut {NoStop}%
\bibitem [{\citenamefont {Hosaka}\ \emph {et~al.}(2005)\citenamefont {Hosaka},
  \citenamefont {Webster}, \citenamefont {Blythe}, \citenamefont {Stannard},
  \citenamefont {Beaton}, \citenamefont {Margolis}, \citenamefont {Lea},\ and\
  \citenamefont {Gill}}]{hosaka2005}%
  \BibitemOpen
  \bibfield  {author} {\bibinfo {author} {\bibfnamefont {K.}~\bibnamefont
  {Hosaka}}, \bibinfo {author} {\bibfnamefont {S.~A.}\ \bibnamefont {Webster}},
  \bibinfo {author} {\bibfnamefont {P.~J.}\ \bibnamefont {Blythe}}, \bibinfo
  {author} {\bibfnamefont {A.}~\bibnamefont {Stannard}}, \bibinfo {author}
  {\bibfnamefont {D.}~\bibnamefont {Beaton}}, \bibinfo {author} {\bibfnamefont
  {H.~S.}\ \bibnamefont {Margolis}}, \bibinfo {author} {\bibfnamefont {S.~N.}\
  \bibnamefont {Lea}}, \ and\ \bibinfo {author} {\bibfnamefont
  {P.}~\bibnamefont {Gill}},\ }\href@noop {} {\bibfield  {journal} {\bibinfo
  {journal} {IEEE Trans. Instr. Meas.}\ }\textbf {\bibinfo {volume} {54}},\
  \bibinfo {pages} {759} (\bibinfo {year} {2005})}\BibitemShut {NoStop}%
\bibitem [{\citenamefont {Balzer}\ \emph {et~al.}(2006)\citenamefont {Balzer},
  \citenamefont {Braun}, \citenamefont {Hannemann}, \citenamefont {Paape},
  \citenamefont {Ettler}, \citenamefont {Neuhauser},\ and\ \citenamefont
  {Wunderlich}}]{balzer2006}%
  \BibitemOpen
  \bibfield  {author} {\bibinfo {author} {\bibfnamefont {C.}~\bibnamefont
  {Balzer}}, \bibinfo {author} {\bibfnamefont {A.}~\bibnamefont {Braun}},
  \bibinfo {author} {\bibfnamefont {T.}~\bibnamefont {Hannemann}}, \bibinfo
  {author} {\bibfnamefont {C.}~\bibnamefont {Paape}}, \bibinfo {author}
  {\bibfnamefont {M.}~\bibnamefont {Ettler}}, \bibinfo {author} {\bibfnamefont
  {W.}~\bibnamefont {Neuhauser}}, \ and\ \bibinfo {author} {\bibfnamefont
  {C.}~\bibnamefont {Wunderlich}},\ }\href@noop {} {\bibfield  {journal}
  {\bibinfo  {journal} {Phys. Rev. A}\ }\textbf {\bibinfo {volume} {73}},\
  \bibinfo {pages} {041407} (\bibinfo {year} {2006})}\BibitemShut {NoStop}%
\bibitem [{\citenamefont {Fawcett}\ and\ \citenamefont
  {Wilson}(1991)}]{Fawcett_AtDatNucDatTab_47_241}%
  \BibitemOpen
  \bibfield  {author} {\bibinfo {author} {\bibfnamefont {B.}~\bibnamefont
  {Fawcett}}\ and\ \bibinfo {author} {\bibfnamefont {M.}~\bibnamefont
  {Wilson}},\ }\href {\doibase http://dx.doi.org/10.1016/0092-640X(91)90003-M}
  {\bibfield  {journal} {\bibinfo  {journal} {Atomic Data and Nuclear Data
  Tables}\ }\textbf {\bibinfo {volume} {47}},\ \bibinfo {pages} {241 }
  (\bibinfo {year} {1991})}\BibitemShut {NoStop}%
\bibitem [{\citenamefont {Harlander}\ \emph {et~al.}(2010)\citenamefont
  {Harlander}, \citenamefont {Brownnutt}, \citenamefont {H{\"a}nsel},\ and\
  \citenamefont {Blatt}}]{Harlander_NJP_12_093035}%
  \BibitemOpen
  \bibfield  {author} {\bibinfo {author} {\bibfnamefont {M.}~\bibnamefont
  {Harlander}}, \bibinfo {author} {\bibfnamefont {M.}~\bibnamefont
  {Brownnutt}}, \bibinfo {author} {\bibfnamefont {W.}~\bibnamefont
  {H{\"a}nsel}}, \ and\ \bibinfo {author} {\bibfnamefont {R.}~\bibnamefont
  {Blatt}},\ }\href@noop {} {\bibfield  {journal} {\bibinfo  {journal} {New
  Journal of Physics}\ }\textbf {\bibinfo {volume} {12}},\ \bibinfo {pages}
  {093035} (\bibinfo {year} {2010})}\BibitemShut {NoStop}%
\bibitem [{\citenamefont {Naegerl}\ \emph {et~al.}(1998)\citenamefont
  {Naegerl}, \citenamefont {Blatt}, \citenamefont {Eschner}, \citenamefont
  {Schmidt-Kaler},\ and\ \citenamefont {Leibfried}}]{naegerl1998coherent}%
  \BibitemOpen
  \bibfield  {author} {\bibinfo {author} {\bibfnamefont {H.}~\bibnamefont
  {Naegerl}}, \bibinfo {author} {\bibfnamefont {R.}~\bibnamefont {Blatt}},
  \bibinfo {author} {\bibfnamefont {J.}~\bibnamefont {Eschner}}, \bibinfo
  {author} {\bibfnamefont {F.}~\bibnamefont {Schmidt-Kaler}}, \ and\ \bibinfo
  {author} {\bibfnamefont {D.}~\bibnamefont {Leibfried}},\ }\href@noop {}
  {\bibfield  {journal} {\bibinfo  {journal} {Optics Express}\ }\textbf
  {\bibinfo {volume} {3}},\ \bibinfo {pages} {89} (\bibinfo {year}
  {1998})}\BibitemShut {NoStop}%
\end{thebibliography}
\end{document}